\documentclass[pra,aps,superscriptaddress,twocolumn,nopacs,nofootinbib]{revtex4}
\usepackage{graphicx,color,epstopdf}
\usepackage{amsmath,amsfonts,enumerate,amsthm,amssymb,mathtools}
\usepackage{enumitem}
\usepackage{thmtools,thm-restate}
\usepackage{bbold}
\usepackage{hyperref}
\usepackage{multirow}
\usepackage{subfigure} % Include figure files
\usepackage{mathdots} %to use futher dots commands

\usepackage{array}
\usepackage{blkarray}

\hypersetup{
colorlinks=true, % false: boxed links; true: colored links, false is default
linkcolor=blue, % color of internal links, red is default
citecolor=blue % color of links to bibliography
}

\usepackage[normalem]{ulem}

\DeclareMathOperator{\Tr}{Tr}

\makeatletter
\def\slashedarrowfill@#1#2#3#4#5{%
  $\m@th\thickmuskip0mu\medmuskip\thickmuskip\thinmuskip\thickmuskip
   \relax#5#1\mkern-7mu%
   \cleaders\hbox{$#5\mkern-2mu#2\mkern-2mu$}\hfill
   \mathclap{#3}\mathclap{#2}%
   \cleaders\hbox{$#5\mkern-2mu#2\mkern-2mu$}\hfill
   \mkern-7mu#4$%
}

\def\rightslashedarrowfilla@{%
  \slashedarrowfill@\relbar\relbar{\raisebox{1.2pt}{$\scriptscriptstyle\diagup$}}\rightarrow}
\newcommand\xslashedrightarrowa[2][]{%
  \ext@arrow 0055{\rightslashedarrowfilla@}{#1}{#2}}

\renewcommand{\v}[1]{\ensuremath{\boldsymbol #1}}

\theoremstyle{plain}

\newtheorem{prop}{Proposition}

\theoremstyle{definition}
\newtheorem{defn}{Definition}
\theoremstyle{remark}

\newtheorem{con}{Conjecture}

\begin{document}

\title{Decomposability and Convex Structure of Thermal Processes}
\begin{abstract}
\end{abstract}

\author{Pawe\l{} Mazurek}
\affiliation{Institute of Theoretical Physics and Astrophysics, National Quantum Information Centre, Faculty of Mathematics, Physics and Informatics, University of Gda\'nsk, 80-308 Gda\'nsk, Poland}

\author{Micha\l{} Horodecki}
\affiliation{Institute of Theoretical Physics and Astrophysics, National Quantum Information Centre, Faculty of Mathematics, Physics and Informatics, University of Gda\'nsk, 80-308 Gda\'nsk, Poland}

\begin{abstract}
We present an example of a Thermal Process for a system of $d$ energy levels, which cannot be performed without an instant access to the whole energy space. This Thermal Process is uniquely connected with a transition between some states of the system, that cannot be performed without access to the whole energy space even when approximate transitions are allowed. Pursuing the question about the decomposability of Thermal Processes into convex combinations of compositions of processes acting non-trivially on smaller subspaces, we investigate transitions within the subspace of states diagonal in the energy basis. For three level systems, we determine the set of extremal points of these operations, as well as the minimal set of operations needed to perform an arbitrary Thermal Process, and connect the set of Thermal Processes with thermomajorization criterion. We show that the structure of the set depends on temperature, which is associated with the fact that Thermal Processes cannot increase deterministically extractable work from a state -- the conclusion that holds for arbitrary $d$ level system. We also connect the decomposability problem with detailed balance symmetry of an extremal Thermal Processes. 
\end{abstract}

\maketitle

\section{Introduction}
One of aims of quantum thermodynamics is to provide such a description of quantum systems interacting with environment that would enable assessment of their usefulness for tasks such as work extraction. Therefore, a question about possible transitions between quantum states, and their energy cost, lies in the center of interest. This question can be posed at a general, model-independent level, when we neglect a precise structure of the system-environment interactions in favor of more general assumptions we impose on them (e. g. energy conservation), and aim at obtaining bounds imposed by quantum mechanics on the performance of quantum systems under these restrictions. 

One of these generalized approaches can be expressed in the language of the resource theory of Thermal Operations \cite{Janzing00} (see also \cite{Streater95}), where, apart from the assumption about the commutation of system-environment interactions with local Hamiltonians, we allow for free addition and erasure of environment state in equilibrium. When restricted to transitions between states diagonal in the basis of a local Hamiltonian, the allowed transformations are described by Thermal Processes -- left stochastic matrices that preserve a Gibbs state. They act on vectors storing states eigenvalues. Thermomajorization criterion \cite{Ruch76, Janzing00,Horodecki13} brings an answer to the question about which states can be achieved from a given initial state under these assumptions and with defined amount of work. 

The thermodynamical description of quantum diagonal states within the resource theory of Thermal Processes has appealing simplicity. 
However, a priori implementation of Thermal Processes requires access to an entire environment. 
Therefore, apart from unitarity and energy conservation, the only thermodynamically-motivated restriction is that the state of environment is a Gibbs one. Such an approach is clearly suitable to 
derive ultimate bounds, however it might be questionable of whether it can be called 
thermodynamics, since the latter not only 
poses limitations on efficiencies of heat engines, but also allows to achieve these limitations (at least in theory) with 
coarse grained operations, that refer only to several relevant macroscopic parameters, such as temperature or pressure. 

Nevertheless, quite recently it was shown that the resource theory of Thermal Processes is indeed thermodynamics in the latter sense. 
Namely, in \cite{Perry16} it was proved that, for diagonal states, all transitions allowed by Thermal Processes can be obtained by 
having microscopic access only to a single qubit of the heat bath. The rest of the bath serves only for simple partial thermalization 
processes which require just weak coupling 
between bath and the system \cite{Alicki07}. This is combined with changing of the Hamiltonian of the system. Thus, while the system (and the single additional qubit of the bath) have to be manipulated microscopically, the heat bath is treated just as in traditional thermodynamics. 
The proposed class of operations (called in \cite{Perry16} "coarse operations"), while fundamentally simple, 
may still be not optimal in practice. In particular, some processes on a single qubit system 
require quite a nontrivial sequence of manipulations on two qubits. 

In contrast, in \cite{Lostaglio16} M. Lostaglio proposes a straightforward implementation of qubit Thermal Processes by considering coupling of a system to a bath via Jaynes-Cummings interaction, and poses the question to what extent qubit Thermal Processes can be universal, i.e. whether a Thermal Process (TP) on higher-dimensional system can be decomposed into a convex combinations of sequences of TPs, where each of the TP acts non-trivially only on a selected pair of the energy levels of the system. This leads to a fundamental problem of specifying some basic TPs, such that:
(i) they can be easily implemented physically, (ii) all transitions allowed by the resource theory of TPs can be obtained from these basic bricks. However, considerations in \cite{Lostaglio16} turn to be based on the assumption of the reversibility of the so-called embedding map \cite{Brandao2015}. This assumption does not hold in general, unless the domain of the map is restricted. Therefore, the question about decomposability of TPs remained open. This assumption was dropped in the recent version of the paper \cite{Lostaglio16b}, published in parallel with this manuscript, and decomposability of TPs into two level TPs was characterized with use of different methods.

In this paper we consider two ways of obtaining all transitions from the basic ones: through compositions of TPs and through convex mixing (possibly interlaced). Our main result is that there is no upper bound on a dimension of the basic bricks,
i.e. for system with $d$ energy levels, there must be a basic operation that involves all $d$ levels. This holds even for approximate transformations, when we allow for the output state to differ from to goal state up to some small value in statistical distance. Note that this result is not 
in contradiction with \cite{Perry16}, where 
thermalizations involve only two levels at a time, because there (unlike here) one also is allowed to change Hamiltonian of the system.

The no-go example for composing TPs out of sequences of TPs acting actively on lower-dimensional subspaces leads to a question about the allowed transitions under operations restricted in this way: What states can be achieved from a given state diagonal in the basis of Hamiltonian of a $d$-level system, if the allowed operations can be composed as mixture of products of Thermal Operations each acting actively on at most $d'$-levels of the system? The second part of this paper is a step to answering this problem by exploring the structure of the set of TPs through calculating and describing properties of extremal points of TPs of $3$-level systems. It enables us to identify all the basic TPs, that allow to obtain arbitrary TP by compositions and mixtures for three level system.

When it comes to answering the above general question, the structure of $d=3$ TPs suggests properties that, if proved general, may be crucial of determining the geometry of the set of TPs for arbitrary $d$, and identifying transitions allowed under the above-mentioned restrictions. Namely, for three level systems, one can determine all extremal TPs using a simple geometrical construction. Furthermore, the geometry of the set of TPs for three level systems changes at the single threshold temperature, where some of the extremal TPs cease to exist. We prove that this property is closely related with the prohibition of increasing deterministically extractable work from the system under TPs, and provide formulas determining values of threshold temperatures for arbitrary $d$-level systems. Finally, we show that the structure of the set of extremal points of TPs might be highly simplified by the symmetry associated with the detailed-balance condition. Namely, we conjecture that every TP that is not self-dual with respect to this symmetry and that is not representable as a simple sum of TPs from subspaces of lower dimensions, cannot be expressed as a mixture of compositions of TPs from these subspaces.

\section{Preliminaria}

We start with characterization of processes that describe transitions between states of system $S$ with fixed Hamiltonian $H_{S}$, that result from its interaction with bath $R$. Later, we will be interested in restrictions on allowed transitions between states of the system, which arise due to limitations we impose on the number of levels of the system that these processes can act actively on.

The interaction with environment is modeled by Thermal Operations. We consider system and bath with respective Hamiltonians $H_{S}=\sum_{i=0}^{d-1}E_{i}|E_{i}\rangle\langle E_{i}|$ and $H_{B}$. 
We denote Gibbs states of the heat bath and the system by 
$\rho_{\beta}^B=e^{-\beta H_{B}}/\Tr[e^{-\beta H_{B}}]$ and  $\rho_{\beta}^S=e^{-\beta H_{S}}/\Tr[e^{-\beta H_{S}}]$,
where $\beta=\frac{1}{kT}$, where $k$ is Boltzmann constant, and $T$ is temperature. 
We now consider the following operations: we can apply to the initial state of the system $\rho_S$ and the Gibbs state of the 
heat bath $\rho_\beta^B$  an arbitrary unitary $U$ which conserves the total energy: $[U,H_{S}+H_{B}]=0$, and then trace out the bath. 
We obtain a trace preserving, completely-positive map on a system $\mathcal{E}(\rho)=\Tr_{B}\big[U\big(\rho\otimes \rho_{\beta}\big)U^{\dagger}\big]$, where $\Tr_{B}$ denotes partial trace over the environment. 

It is visible that the map preserves the Gibbs state $\rho_{\beta}^{S}$. From the assumption of energy conservation it follows that elements $\rho^{(\omega)}$ of the matrix $\rho=\sum_{\omega}\rho^{(\omega)}$, such that $\rho^{(\omega)}=\sum_{n,m:E_{n}-E_{m}=\omega}\rho_{n,m}|E_{n}\rangle\langle E_{m}|$, are transformed independently: $\mathcal{E}(\rho^{(\omega)})=\mathcal{E}(\rho)^{(\omega)}$ \cite{Lostaglio15} (see also \cite{Cwiklinski2015}). In particular, it shows that if one starts with $\rho$ such that $[\rho,H_{S}]=0$ (no coherences in the eigenbasis of the system Hamiltonian), one cannot obtain coherences through Thermal Operations. Therefore, we define the basic object of interest of the paper: 

\begin{defn}
Take states $\rho$ and $\sigma$ such that $[\rho,H_{s}]=[\sigma,H_{s}]=0$, and eigenvalues in the eigenbasis of $H_{S}$ of these vectors are represented by vectors $\v{p}$ and $\v{r}$, respectively. 
A Thermal Process is a stochastic map $T$: $T\v{p}=\v{r}$ that corresponds to a Thermal Operation $\mathcal{E}(\rho)=\sigma$. 
\end{defn}

From above it is visible that every TP can be represented as a left stochastic (i.e., with elements summing to $1$ within each column), Gibbs preserving matrix $T$: $T\v{g}=\v{g}$, where $\v{g}$: $g_{i}=q_{i,0}/\sum\limits_{j} q_{j,0}$. Without loss of generality here and in the whole paper we assume that the ground state energy of the system is zero: $E_{0}=0$. We index rows and columns of matrices from $0$ to $d-1$. We will also use a shorthand notation $q_{m,n}=e^{-\beta(E_{m}-E_{n})}$. Conversely, every left stochastic, Gibbs preserving matrix leads to a Thermal Operation on a diagonal state \cite{Horodecki13}. Therefore, the set of TPs and a set of left stochastic, Gibbs preserving matrices are equal, and we focus on the latter. 

\section{A non-decomposable Thermal Process in an arbitrary dimension}\label{transition}

Below we show that for a $d$ level system, one can always find a pair of states $\v{p}$ and $\v{r}$ such that they are connected by a TP $P(\v{p})=\v{r}$, and such that there is no other process connecting the states, and $P$ cannot be decomposed into a convex combination of compositions of Thermal Processes, each acting on at most $d-1$ dimensional subspaces.

In Section \ref{approx} we show that a state $\v{r}$ cannot be achieved by 2 level TPs from $\v{p}$ even approximatively: there exists $\epsilon>0$ such that all states $\v{r}'$ achievable from $\v{p}$ by 2 level TPs satisfy $||\v{r}-\v{r}'||\geq \epsilon$.

We take 
\begin{eqnarray}\label{states}
\v{p}=
\begin{pmatrix}
1\\
0\\
0\\
\dots\\
0
\end{pmatrix},
&
\v{r}=
\begin{pmatrix}
1-\sum\limits_{i=1}^{d-1}q_{i,0}\\
q_{1,0}\\
q_{2,0}\\
\dots\\
q_{d-1,0}
\end{pmatrix}.
\end{eqnarray}

Note that, in order to assure that $\v{r}$ represents a state, we have to assume $\sum\limits_{i=1}^{d-1}q_{i,0}\leq 1$. One can always find temperature low enough such that the above is satisfied. In the following section we will provide examples of non-decomposable transitions for higher temperatures. Note also that $\v{r}$ does not represent a Gibbs state $\v{g}$: $g_{i}=q_{i,0}/\sum\limits_{j} q_{j,0}$. Nevertheless, proportions between occupations on levels $1,\dots,d-1$ remain the same as for $\v{g}$.

From $P(\v{p})=\v{r}$ we see that the $0$-th element of the $0$-th row of $P$ (i.e. $P_{0,0}$) is equal to $1-\sum\limits_{i=1}^{d-1}q_{i,0}$ (see (\ref{T})).
The Gibbs preserving condition ($\sum\limits_{j=0}^{d-1}P_{i,j}q_{j,0}=q_{i,0}$) applied to the zeroth row ($i$=0) implies then that other elements of this row are equal to 1 (i.e. $\forall_{j>0} P_{0,j}=1$). In turn, the stochasticity condition ($\forall_{j}\sum\limits_{i=0}^{d-1}P_{i,j}=1$) applied to columns $j>0$ implies then $\forall_{i>0,j>0} P_{i,j}=0$. Then, the Gibbs preserving condition applied to rows $i>0$ uniquely determines $P_{i,0}$, and every TP transforming $\v{p}$ into $\v{r}$ has to take a form
\begin{equation}\label{T}
P=
\begin{pmatrix}
1-\sum\limits_{i=1}^{d-1}q_{i,0} & 1 & 1 & \dots & 1 \\
q_{1,0} & 0 & 0 & \dots & 0 \\
q_{2,0} & 0 & 0 & \dots & 0 \\
\dots & \dots & \dots& \dots& \dots\\
q_{d-1,0} & 0 & 0 & \dots & 0 
\end{pmatrix}.
\end{equation}

Now we show that $P$ cannot be decomposed as a composition of TPs, each acting on at most $d-1$ dimensional subspace. Every such decomposition would take a form $P=AB$, where both $A$ and $B$ are TPs. We will show below that if $A$ and $B$ are left stochastic and Gibbs preserving, then one of the matrices has to be equal to $P$. Therefore, it is impossible to decompose $P$ into two TPs that act non-trivially on at most $d-1$ dimensional subspaces. It follows that the above conclusion holds for a decomposition constructed as a product of an arbitrary natural number of TPs: If it was possible to decompose $P$ into $k$ TPs, each acting on a $d-1$ dimensional subspace, then one could compress $k-1$ of them into a matrix that will be a TP, but from the above we see that it has to be equal either to $P$ or to an operation that acts trivially on $0-th$ level. If it is an operation that acts trivially on $0$-th level, it can be decomposed only to such operations. If it is equal to $P$, we proceed in decomposing it into TPs, at every step dividing the decreasing number of processes into two groups: one composed of one TP, and the other composed of remaining ones. In this way we see that for arbitrary natural $k$, for every decomposition of $P$ into $k$ TPs, it has to be of the form $X_{1}\dots X_{m}P Y_{1}\dots Y_{n}$, where TPs $X_{i}$, $Y_{j}$, for $i=1,\dots,m$, $j=1,\dots,n$ and $n+m=k-1$, act trivially on the $0$-th level. 

We begin to show that a decomposition $P=AB$ leads to one matrix that acts trivially on $0$-th level and one that is equal to $P$. Let us notice that the condition $\forall_{i>0,j>0} P_{i,j}=0$ implies that the product of an $i$-th row ($i>0$) of $A$ and $j$-th column ($j>0$) of B has to be zero. As these matrices can store only non-negative entries, this implies that $\forall_{i>0,j>0} A_{i,0}B_{0,j}=0$. Assume now that there is some $k>0$ such that $B_{0,k}\neq 0$. This implies $\forall_{i>0}A_{i,0}=0$ so that $\forall_{i>0} A_{i,0}B_{0,k}=0$ can be fulfilled. But then, from the stochasticity condition applied to the $0$-th column of $A$ we have $A_{0,0}=1$, and, from Gibbs preserving condition applied to the $0$-th row of A, $\forall_{j>0}A_{0,j}=0$. As we already saw before, this implies $B_{0,0}=1-\sum\limits_{i=1}^{d-1}e^{-\beta\Delta_{i0}}$, which enforces $B=P$, and leads to the thesis.
On the other hand, if there is no $k>0$ such that $B_{0,k}\neq 0$, then, from Gibbs preserving condition applied to the $0$-th row of $B$, we have $B_{0,0}=1$, which implies $\forall_{i>0}B_{i,0}=0$ from stochasticity condition applied to the first column of $B$. In order to have $P=AB$, we must then have $A_{0,0}=1-\sum\limits_{i=1}^{d-1}e^{-\beta\Delta_{i0}}$, which implies $A=P$.

Finally we will show that $P$ is an extreme point of TPs, and therefore, cannot be formed as a convex combination of other TPs. The set of TPs is convex, which follows from its equivalence to the set of left stochastic, Gibbs-preserving matrices -- both stochasticity and Gibbs-preserving properties are linear. One can easily show that, in a case of $d\times d$ TP, there are always $2d-1$ linearly independent restrictions on this process (arising from $d$ stochastic conditions applied to the columns and $d$ Gibbs-preserving conditions applied to the rows). As every linearly-independent restriction applied to the set of matrices can only increase by 1 a number of non-zero elements in every extremal point of the set of such matrices, every TP with less than $d^2-(2d-1)=(d-1)^2$ zero elements is not an extreme point of the set \cite{Barrett05}. One can therefore construct the set of all extremal points of TPs by fixing $(d-1)^2$ elements to be zero, and continue fixing to zero more elements until the remaining ones are fixed by stochasticity and Gibbs-preserving conditions -- a sign that the corresponding processes cannot be decomposed into a sum of other processes with at least $(d-1)^2$ zero elements. As fixing $P_{i,j}=0$ $\forall_{i>0,j>0}$ implies values of the first row and the first column of $P$, it shows that $P$ is an extremal point of the set of TPs for $d$ level system.

Again, let us stress that the condition for all the elements of the matrix $P$ to be non-negative implies that temperature has to be low enough to ensure $\sum\limits_{i=1}^{d-1}q_{i,0}\leq 1$.

\section{Structure of the set of extremal Thermal Processes}
By following the procedure outlined above, one can, in principle, find all extremal points of TPs for arbitrary dimension $d$. These extremal points will be further denoted as EPTP(d).  Alternatively, one can apply a procedure of generating the whole set of extremal points from a trivial extremal point (identity), presented in \cite{Gregory92}. In any case, obtaining this set explicitly is demanding for increasing local dimension $d$. At the end of this section, we point out a property of extremal points of TPs for three level systems that, if it holds for arbitrary $d$, would provide an intuitive, graphical way of obtaining extremal points of $d$ level TPs.\\

For a 2 level system, the structure of the set is straightforward, with only two extremal points: 
\begin{equation}\label{2levels}
\text{EPTP(2)}=\left \{{Id(2)}, 
\begin{pmatrix}
1-q_{1,0}& 1 \\
q_{1,0} & 0 
\end{pmatrix}
\right\}, 
\end{equation}
where by $Id(d)$ we denote the identity $d\times d$ matrix.

\begin{figure*}
\centering
\includegraphics[width = 1.0\linewidth]{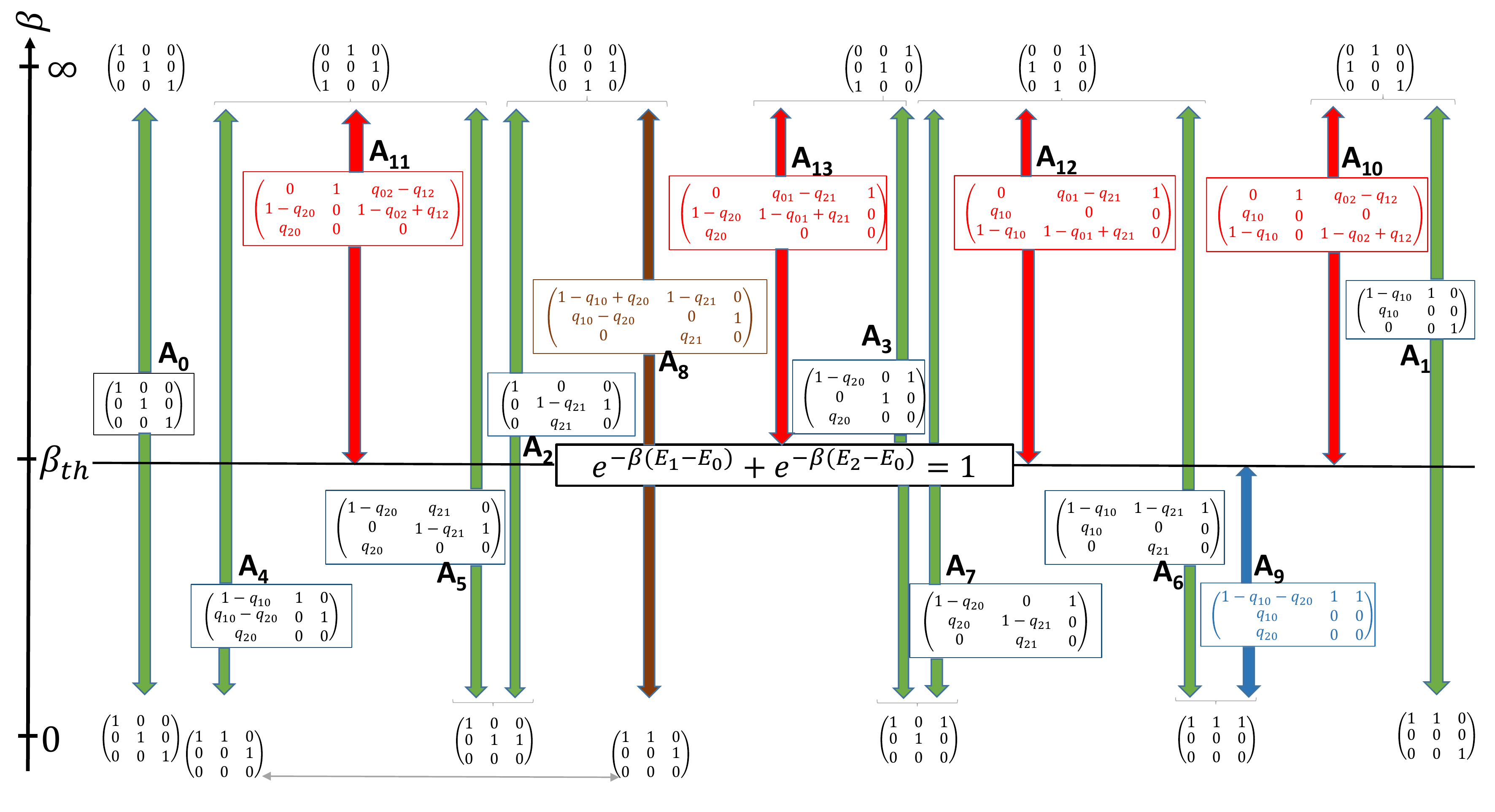}
\caption{\label{fig:sets} Presence of extremal points of TPs for $d=3$ level systems, indicated by arrows, for different temperatures (vertical axis). Some extremal points exist only in a selected temperature range. In zero and infinite temperatures, some processes coincide, which is indicated by connecting gray horizontal arrows and braces. In infinite temperatures ($\beta=0$), extremal points are permutation matrices, in accordance with Birkhoff theorem \cite{Birkhoff46}. All extremal points decomposable into a product of extremal points acting non-trivially on at most 2 levels are represented by green arrows. Red arrows are associated with processes valid below the threshold temperature, and a blue arrow corresponds to a process above this temperature. Brown color distinguishes a non-decomposable process present in the whole temperature range.}\label{Fig1}
\end{figure*}

\subsection{Structure of the set for $d=3$ level systems.}

For a $d=3$ level system, the geometry of the set becomes temperature dependent (see Fig. \ref{Fig1}). Below a threshold temperature 
$T_0=1/k\beta_0$ defined by the following relation 
\begin{equation}
\label{eq:t_crit}
q_{1,0}+q_{2,0}=e^{-\beta_0 E_{1}}+e^{-\beta_0 E_{2}}=1,
\end{equation}
it can be expressed as 
\begin{equation}
\text{EPTP(3)}^{\beta\geq\beta_{0}}=\text{EPTP(3)}^{univ}\cup\{A_{9}\},
\end{equation}
whereas in the remaining regime
\begin{equation}
\text{EPTP(3)}^{\beta\leq\beta_{0}}=\text{EPTP(3)}^{univ}\cup\{A_{10},A_{11},A_{12},A_{13}\}.
\end{equation}

The set 
\begin{equation}
\text{EPTP(3)}^{univ}=\{A_{0},A_{1},A_{2},A_{3},A_{4},A_{5},A_{6},A_{7},A_{8}\}
\end{equation}
of extremal points that are present for the whole spectrum of temperatures contains an identity matrix $A_{0}=Id(3)$ and two-level Thermal
Processes ($A_{1}$, $A_{2}$ and $A_{3}$):

\begin{equation}\label{A6}
A_{1} = \begin{pmatrix}
1-q_{1,0}& 1 & 0 \\
q_{1,0} & 0 & 0 \\
0 & 0 & 1 
\end{pmatrix},
\end{equation}

\begin{equation}\label{A7}
A_{2} = \begin{pmatrix}
1 & 0 & 0 \\
0 & 1-q_{2,1} & 1 \\
0 & q_{2,1} & 0 
\end{pmatrix},
\end{equation}

\begin{equation}\label{A8}
A_{3} = \begin{pmatrix}
1-q_{2,0}& 0 & 1 \\
0 & 1 & 0 \\
q_{2,0} & 0 & 0 
\end{pmatrix},
\end{equation} 

apart from extremal processes that can be expressed as products of two-level processes:

\begin{equation}\label{A1}
A_{4} = A_{2}A_{1}=\begin{pmatrix}
1-q_{1,0}& 1 & 0 \\
q_{1,0}-q_{2,0} & 0 & 1 \\
q_{2,0} & 0 & 0 
\end{pmatrix},
\end{equation}

\begin{equation}\label{A2}
A_{5} = A_{3}A_{2}=\begin{pmatrix}
1-q_{2,0}& q_{2,1} & 0 \\
0 & 1-q_{2,1} & 1 \\
q_{2,0} & 0 & 0 
\end{pmatrix},
\end{equation}

\begin{equation}\label{A3}
A_{6} = A_{1}A_{2}=\begin{pmatrix}
1-q_{1,0}& 1-q_{2,1} & 1 \\
q_{1,0} & 0 & 0 \\
0 & q_{2,1} & 0 
\end{pmatrix},
\end{equation}

\begin{equation}\label{A5}
A_{7} = A_{2}A_{3}=\begin{pmatrix}
1-q_{2,0}& 0 & 1 \\
q_{2,0} & 1-q_{2,1} & 0 \\
0 & q_{2,1} & 0 
\end{pmatrix}.
\end{equation}

The last member of $EPTP(3)^{univ}$ cannot be expressed in such a way:
\begin{equation}\label{A0}
A_{8} = 
\begin{pmatrix}
1-q_{1,0}+q_{2,0} & 1-q_{2,1} & 0 \\
q_{1,0}-q_{2,0} & 0 & 1 \\
0 & q_{2,1} & 0 
\end{pmatrix}.\\\end{equation}

The remaining extremal points are present only in temperatures higher or lower than the threshold temperature $T_{0}$ of \eqref{eq:t_crit}, 
which is associated with the requirement, coming from stochasticity of the matrices, that all of their elements take values from 
a range $[0,1]$. For temperature above the $T_{0}$ we have four extremal points: 

\begin{equation}\label{A0}
A_{10} = 
\begin{pmatrix}
0 & 1 & q_{0,2}-q_{1,2} \\
q_{1,0} & 0 & 0 \\
1-q_{1,0} & 0 & 1-q_{0,2}+q_{1,2} 
\end{pmatrix}.\\\end{equation}
\begin{equation}\label{A0}
A_{11} = 
\begin{pmatrix}
0 & 1 & q_{0,2}-q_{1,2} \\
1-q_{2,0} & 0 & 1-q_{0,2}+q_{1,2} \\
q_{2,0} & 0 & 0 
\end{pmatrix}.\\\end{equation}
\begin{equation}\label{A0}
A_{12} = 
\begin{pmatrix}
0 & q_{0,1}-q_{2,1} & 1 \\
q_{1,0} & 0 & 0 \\
1-q_{1,0} & 1-q_{0,1}+q_{2,1} & 0 
\end{pmatrix}.\\\end{equation}
\begin{equation}\label{A0}
A_{13} = 
\begin{pmatrix}
0 & q_{0,1}-q_{2,1} & 1 \\
1-q_{2,0} & 1-q_{0,1}+q_{2,1} & 0 \\
q_{2,0} & 0 & 0 
\end{pmatrix}.\\\end{equation}
Below threshold temperature, all the above four points disappear, and instead a single extremal point emerges, which is the map $P$ 
from the previous section:

\begin{equation}
A_{9} = P(3)=\begin{pmatrix}
1-q_{1,0}-q_{2,0}& 1 & 1 \\
q_{1,0} & 0 & 0 \\
q_{2,0} & 0 & 0 
\end{pmatrix},
\end{equation}

We have already shown that $A_{9}$ cannot be decomposed to a product of two level TPs, and that there exist states $\v{p}$ and $\v{r}$ such that $A_{9}\v{p}=\v{r}$. The same remains true for maps $A_{10},A_{11},A_{12}$ and $A_{13}$: if, for $\beta\leq \beta_{0}$ and an arbitrary $0\leq a\leq1$, one takes

\begin{eqnarray}\nonumber
\v{p}=
\begin{pmatrix}
1\\
0\\
0
\end{pmatrix},
&
\v{r}=
\begin{pmatrix}
0\\
a\\
1-a\\
\end{pmatrix},
\end{eqnarray}
it is clear that the only TP $R$ satisfying $R(\v{p})=\v{r}$, has to have $R_{0,0}$=0, and therefore is a convex combination of $A_{10}-A_{13}$. As no such a process can be constructed as a product of two-level TPs ($A_{1}-A_{3}$), also these TPs lead to an example of operations allowed by Thermal Operations Resource Theory, that cannot be performed as a convex combination of processes that act non-trivially only on pairs of energy levels. 

Therefore, we arrive at  
\begin{prop}
For a 3 level diagonal system, the set of operations that, by mixtures and compositions, enables to perform an arbitrary transformation allowed by Thermal Operations, is $\{Id(3), A_{1},A_{2},A_{3},A_{8},A_{9}\}$ for temperatures that satisfy $e^{-\beta E_{1}}+e^{-\beta E_{2}}\leq 1$, and $\{Id(3), A_{1},A_{2},A_{3},A_{8},A_{10},A_{11},A_{12},A_{13}\}$ for temperatures $e^{-\beta E_{1}}+e^{-\beta E_{2}}\geq 1$.
\end{prop}

\subsection{Detailed balance symmetry.}
\begin{figure*}
\centering
\includegraphics[width = 1.0\linewidth]{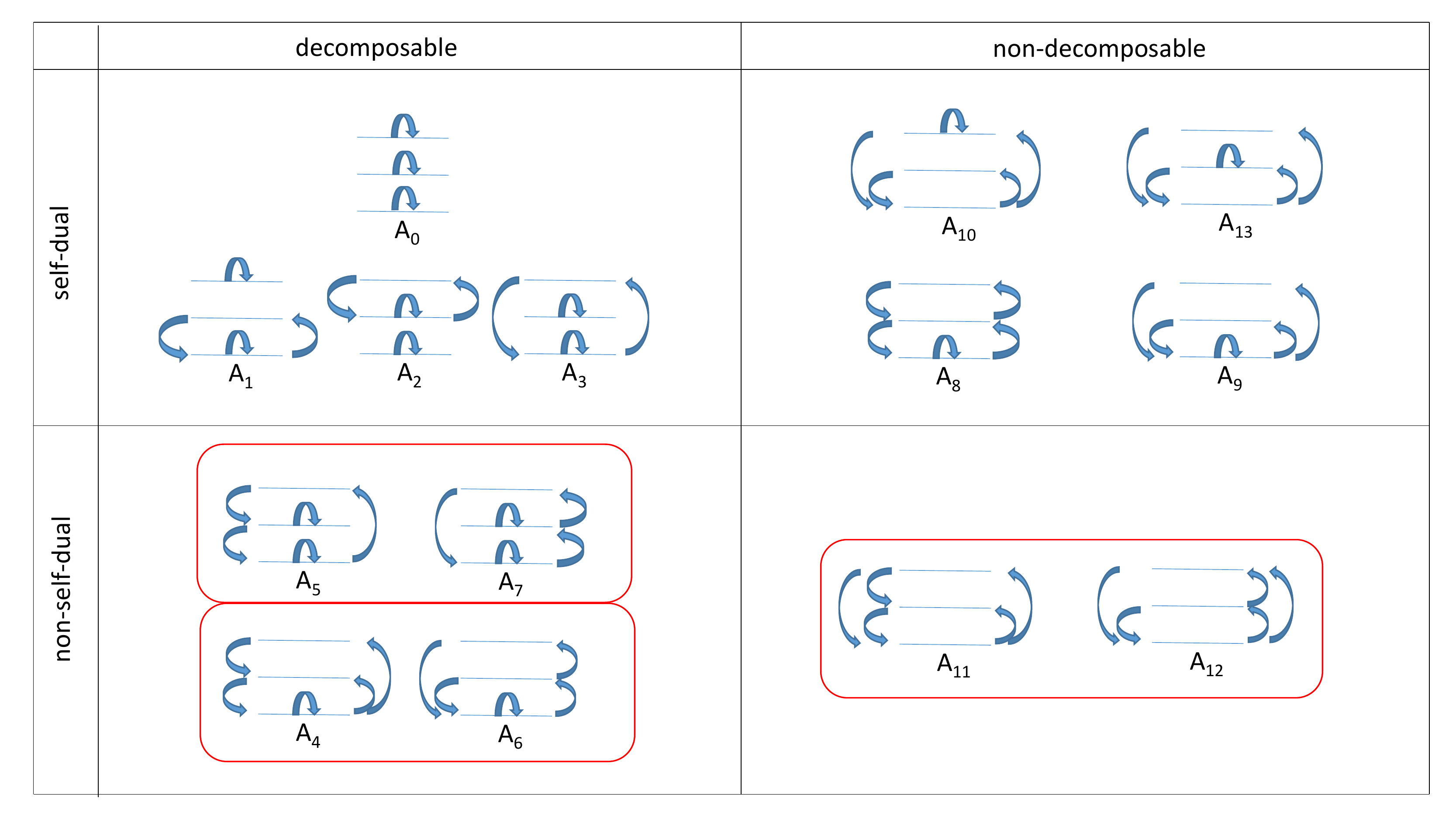}
\caption{\label{fig:symmetry} 
Extremal points of TPs for three level systems. Arrows indicate a transformation between selected levels occurring with non-zero probability. A conjugation $A_{i}\rightarrow \tilde{A}_{i}$ is equivalent to redirecting arrows. Pairs of extremal points connected via the conjugation are contained within red frames. Note that none of the non-trivial self-dual three level extremal processes is decomposable into a sequence of three level processes.}
\label{symmetry}
\end{figure*}

Below we point out a symmetry of extermal points of TPs that can be associated with detailed-balance condition. For a system with a Hamiltonian $H$, let us define a scalar product $\langle X|Y \rangle_\beta$ between two observables $X$ and $Y$ by $\langle X|Y \rangle_\beta=\Tr[X Y^{\dagger} \rho_{\beta}]$, with a Gibbs state $\rho_{\beta}=e^{-\beta H}/ \Tr[e^{-\beta H}]$. One defines a conjugate of an operator with respect to this scalar product:
$\langle \tilde{A}(X)|Y\rangle_\beta=\langle X|A(Y)\rangle_\beta$. 
It follows that $\tilde{A}_{i,j}=A_{j,i}^{T}e^{-\beta(E_{i}-E_{j})}$. It can be rewritten as $\tilde{A}=M_{\rho_{\beta}}A^{T}M_{\rho_{\beta}}^{-1}$, where $M_{\rho_{\beta}}=diag[1,q_{1,0},\dots,q_{d-1,0}]$ is a matrix storing on its diagonal values proportional to occupations in a Gibbs state. 
Self-duality with respect to such a scalar product ($\tilde{A}=A$) served as a definition of detailed balance for generator of 
dynamical semigroup \cite{Alicki76, Kossakowski77,Alicki07}. 
The conjugation is linear and maps left stochastic and Gibbs preserving maps into themselves, and conserves the number of non-zero elements in their matrix representation. Furthermore, as the conjugation is its inverse, the orbits of maps associated with the conjugation are composed only of 1 or 2 elements. Therefore, all extremal points of TPs are mapped to extremal points of TPs. If it was not true, then we can could write $\tilde{A}=\lambda \tilde{A}_{1} + (1-\lambda) \tilde{A}_{2}$ for $\tilde{A}_{1}\neq \tilde{A}_{2}$, $0<\lambda<1$ and some extremal $A$, from which we would have $A=\lambda A_{1} + (1-\lambda) A_{2}$, which contradicts the thesis that $A$ is extremal (as $\tilde{A}_{1}\neq \tilde{A}_{2}$ implies $A_{1}\neq A_{2}$).

Below we describe dual properties of TPs for the case $d=3$. We see that extremal points of TPs from the set
$\{A_{1}, A_{2}, A_{3},A_{8}, A_{9}, A_{10}, A_{13}\}$ are self-dual with respect to this conjugation, while 
$(A_{4}, A_{6})$, $(A_{5}, A_{7})$, $(A_{11}, A_{12})$ form pairs of extremal points of which one element is a conjugate of another. From the physical point of view, the conjugation of a TP reverses the direction of every transformation between levels of the system that the TP is defining. This is shown in Fig. \ref{fig:symmetry}, where self-dual and non self-dual extremal TPs for three level systems are grouped with respect to the ability of composing them from two level TPs. Note that, among elements that act non-trivially on all levels, there are no extremal TPs that are self-dual and can be decomposed as a sequence of extremal TPs from a lower dimensional space. Therefore we propose the following conjecture: 
\begin{con}
If an extremal TP $C$ for $d$ dimensional space is decomposable into a sequence of extremal TPs $A$, $B$, each acting non-trivially on at most $d-1$ dimensional space: $C=AB$, and $C$ is not a direct sum of extremal TPs from lower dimensional subspaces, then $C$ is not self-dual with respect to the conjugation associated with the operator scalar product.  
\end{con}
Above, by demanding that $C$ be not a direct sum of TPs from lower dimensional spaces, we account for cases of self-dual $A$ and $B$ acting on disjoint subspaces, trivially leading to a self-dual $C$. The main concern in describing the set of TPs for arbitrary $d$ is the construction and characterization of structures that emerge with the increasing space dimension. Proving the above conjecture might be helpful in shedding more light onto this problem.\\
In the next section we present another useful property of extremal TPs for three level system -- their connection to certain type of transformations of curves on the so called thermomajorozation diagrams.

\subsection{Connection to thermomajorization diagrams.}\label{thermo}
The continuity of the transition between $A_{9}$ and $A_{10}-A_{13}$ extremal points at $\beta=\beta_{0}$ is even more visible when one takes into account properties of states that are transformed by these extremal processes. In order to examine this, we invoke the notion of thermal order, associated with thermomajorization criterion.

\begin{defn}[Thermo-majorization curve]
Define a vector $\v{s}=(q_{00},q_{10},q_{20},\dots,q_{d-1,0})$. For every state $\rho$ commuting with $H_{S}$, let a vector $\v{p}$ represents occupations $p_{i}$ of energy levels $E_{i}$, $i=0,1,\dots,d-1$. Choose a permutation $\pi$ on \v{p} and \v{s}, such that it leads to a non-increasing order of elements in a vector $\v{d}$, $d_{k}=\big(\frac{\sum_{i=0}^{k}(\pi \v{p})_i}{ \sum_{i=0}^{k} (\pi \v{s})_i }\big)$, $k=0,\dots,d-1$. A set of points $\{\sum_{i=0}^{k}(\pi\v{p})_i,\sum_{i=0}^{k}(\pi\v{s})_i\}_{k=0}^{d-1}\cup\{0,0\}$, connected by straight lines, defines a curve associated with the state $\rho$. We denote it by $\beta(\v{p})$ and call a thermomajorization curve of state $\rho$ represented by $\v{p}$. 
\end{defn}

Points $\{\sum_{i=0}^{k}(\pi\v{p})_i,\sum_{i=0}^{k}(\pi\v{s})_i\}_{k=0}^{d-1}$ will be called elbows of a curve $\beta(\v{p})$. 
The curve is convex due to a non-increasing order of elements in $\v{d}$. Let us note that there might be more than one permutation leading to a creation of a convex curve $\beta(\v{p})$. The vector $\pi(1,\dots,d)^{T}$ will be called a $\beta$-order of $\v{p}$. It shows modification of the order of segments that had to be done in order to assure convexity of $\beta(\v{p})$.

All transitions between diagonal states under TPs are described by the following criterion:

\begin{prop}\label{term} [\cite{Horodecki13}]
A transition from $\v{p}$ to $\v{r}$ under TPs is possible if and only if $\beta(\v{p})$ thermomajorizes $\beta(\v{r})$, i.e. all elbows of $\beta(\v{r})$ lie on $\beta(\v{p})$ or below it.\\
\end{prop}

From the structure of extremal points of TPs for 2 and 3 level systems, the following Proposition can be shown:

\begin{prop}\label{Prop3}
For every extremal TP $R$ for 2 and 3 level systems, there exists a permutation $\kappa$ such that all states $\v{p}$ with $\beta$-order $\kappa(1,\dots,d)^{T}$ are transformed by $R\v{p}=\v{r}$ into states of the same $\beta$-order $\kappa'(1,\dots,d)^{T}$ . Moreover, all elbows of $\beta(\v{r})$ lie exactly on $\beta(\v{p})$. 
\end{prop}

The proof of the above property for every extremal TP $A$ can be expressed with the help of a matrix that will be denoted $A^{s}$ and that describes the transformation performed by the process $A$ on slopes of the thermomajorization curve of an initial state. 
Below, we show the exact calculations for the case $A=A_{8}$.\\

For a vector $\v{p}$, define an associated vector $\partial \v{p}: \partial p_{i}=p_{i}q_{0,i}$. It represents slopes of segments of $\beta(\v{p})$; $\partial p_{i}$ is a slope of a segment associated with the level $i$, with population $p_{i}$. It can be easily shown that a map $A^{s}$, associated with a map $A\v{p}=\v{r}$ and such that $A^{s}\partial\v{p}=\partial\v{r}$, takes the form $A^{s}=M_{\rho_{\beta}}^{-1}AM_{\rho_{\beta}}$. It satisfies $A^{s}=\tilde{A}^{T}$, and is a counterpart of $A$, in a sense that it satisfies stochasticity condition for every row: 
\begin{flalign}\label{stoch}
&A^{s} \begin{pmatrix}
1\\
\dots\\
1
\end{pmatrix}
=M_{\rho_{\beta}}^{-1}AM_{\rho_{\beta}}\begin{pmatrix}
1\\
\dots\\
1
\end{pmatrix}=\nonumber&\\
&
=
M_{\rho_{\beta}}^{-1}A \begin{pmatrix}
1\\
\dots\\
q_{d-1,0}
\end{pmatrix}
=
M_{\rho_{\beta}}^{-1} \begin{pmatrix}
1\\
\dots\\
q_{d-1,0}
\end{pmatrix}= \begin{pmatrix}
1\\
\dots\\
1
\end{pmatrix},&
\end{flalign}
and Gibbs-preserving condition for every column:

\begin{flalign}\label{Gibbs}
&(A^{s})^{T} \begin{pmatrix}
1\\
\dots\\
q_{d-1,0}
\end{pmatrix}
=M_{\rho_{\beta}}A^{T}M_{\rho_{\beta}}^{-1}\begin{pmatrix}
1\\
\dots\\
q_{d-1,0}
\end{pmatrix}=&\nonumber\\&=
M_{\rho_{\beta}}A^{T} \begin{pmatrix}
1\\
\dots\\
1
\end{pmatrix}
=
M_{\rho_{\beta}}\begin{pmatrix}
1\\
\dots\\
1
\end{pmatrix}= \begin{pmatrix}
1\\
\dots\\
q_{d-1,0}
\end{pmatrix},
\end{flalign}
where third equalities in (\ref{stoch}) and (\ref{Gibbs}) come from (row) Gibbs-preserving and (column) stochasticity of $A$, respectively.

Therefore, the thermal process $A_{8}$:

\begin{flalign}\label{pq}
\v{p}=
\begin{pmatrix}
a\\
b\\
c
\end{pmatrix}
\xrightarrow{A_{8}} 
\v{r}=& \begin{pmatrix}
a(1-q_{1,0}+q_{2,0})+(1-q_{2,1})b\\
a(q_{1,0}-q_{2,0})+c\\
b~ q_{2,1} 
\end{pmatrix}.&
\end{flalign}

is associated with the following transformation of slopes of the segments of $\beta(\v{p})$: 
\begin{flalign}\label{As}
\partial \v{p}=
\begin{pmatrix}
\alpha\\
\gamma\\
\delta
\end{pmatrix}
\xrightarrow{A_{8}^{s}} 
\partial \v{r}=& \begin{pmatrix}
(1-q_{1,0}+q_{2,0})\alpha+(q_{1,0}-q_{2,0})\gamma\\
(1-q_{2,1})\alpha+ q_{2,1}\delta\\
\gamma
\end{pmatrix}. 
\end{flalign}

\begin{table}
\begin{tabular}{|c|c|c|}
\hline
& $\beta$-order  &  extremal point $A_{i}$ \\
& of $\v{p}$  & and  $\beta$-order of $\v{r}=A_{i}\v{p}$\\\cline{1-3}
\multirow{3}{*}{\rotatebox{90}{$\beta\geq\beta_{0}$}} & (312) &  $A_{1}(321),A_{3}(132),A_{4}(231),A_{7}(123),A_{8}(213) $ \\\cline{2-3}
& (321) &  $  A_{1}(312),A_{2}(231),A_{5}(213),\boldsymbol{A_{9}}(132)\lor(123)$ \\\cline{2-3}
& (231) &  $ A_{2}(321),A_{3}(213),A_{6}(312),\boldsymbol{A_{9}}(132)\lor(123)$\\\cline{1-3}
\multirow{3}{*}{\rotatebox{90}{$\beta\leq\beta_{0}$}}& (312)  & $A_{1}(321),A_{3}(132),A_{4}(231),A_{7}(123),A_{8}(213) $ \\\cline{2-3}
& (321) &  $A_{1}(312),A_{2}(231),A_{5}(213),\boldsymbol{A_{12}}(132), \boldsymbol{A_{13}}(123)$ \\\cline{2-3}
& (231) &  $A_{2}(321),A_{3}(213),A_{6}(312),\boldsymbol{A_{10}}(132), \boldsymbol{A_{11}}(123)$\\\cline{1-3}
\end{tabular}
\caption{Extremal points $A_{i}$ that map a state $\v{p}$ with a given $\beta$-order to a state $\v{r}$ with a fixed $\beta$-order, and such that all elbows of $\beta(\v{r})$ lie on $\beta(\v{p})$. Transitions performed by $\boldsymbol{A_{9}}$ in low temperatures can be achieved by $\{\boldsymbol{A_{10}},\boldsymbol{A_{11}},\boldsymbol{A_{12}},\boldsymbol{A_{13}} \}$ in high temperatures. For the transitions performed by $\boldsymbol{A_{9}}$, slopes of the last two segments of $\v{r}$ are equal, henceforth the degeneration of the $\beta$-order. Information about all possible transformations stem from the above table and an observation that reversing $\beta$-order of $\v{p}$ is reflected in the reversed $\beta$-order of $\v{r}$: e.g. for $\v{p}$ with $\beta$-order (213) we obtain, through $A_{1}$, a state $\v{r}$ with $\beta$-order (123).  }\label{Tab22}. 
\end{table}

\begin{figure}[h]
\centering
\includegraphics[width = 1.0\linewidth]{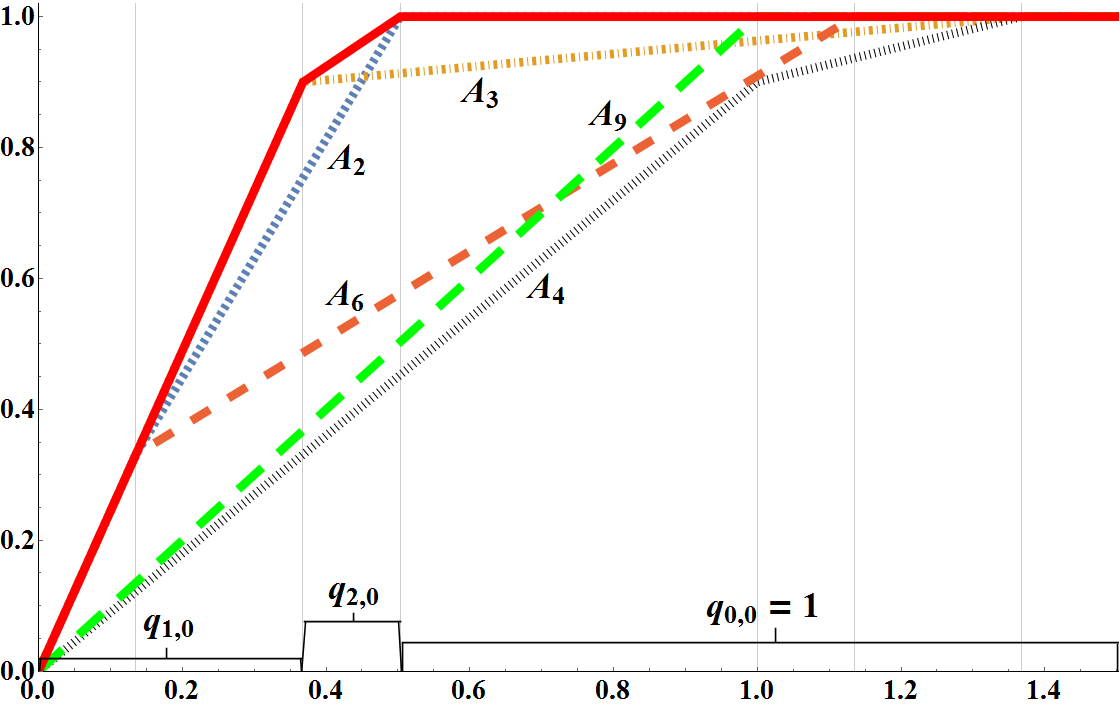}
\caption{\label{fig:sets} 
Thermomajorization curves of a state $\v{p}=(0,0.9,0.1)$ with $\beta$-order (2,3,1) (solid red), and states obtained from it by applying extremal TPs: $A_{2}$ (dashed blue), $A_{3}$ (dot-dashed orange), $A_{4}$ (dotted black), $A_{6}$ (medium-dashed brown), $A_{9}$ (long-dashed green). We have chosen a Hamiltonian such that $\beta E_{1}=1$ and $\beta E_{2}=2,$ which implies $q_{1,0}+q_{2,0}< 1$. i.e. that we are in temperature range $\beta>\beta_{0}$. Curve $\beta(A_{4}\v{p})$ does not have all elbows on the initial curve $\beta({\v{p}})$; $\v{p}$ with another $\beta$-order would be required for $\beta(A_{4}\v{p})$ to have all elbows on $\beta(\v{p})$ (see Table \ref{Tab22}). Due to low temperature regime, degeneration of $\beta$-order of $\beta(A_{9}\v{p})$ occurs -- both vectors (1,2,3) and (1,3,2) are the correct $\beta$-orders (cf. Fig. \ref{Fig2}). 
}
\label{Fig3}
\end{figure}

Let's assume that $\beta$-order of $\v{p}$ is $(3,1,2)$, i.e. $\delta\geq\alpha\geq\gamma$ . As it implies $(1-q_{2,1})\alpha+ q_{2,1}\delta\geq(1-q_{1,0}+q_{2,0})\alpha+(q_{1,0}-q_{2,0})\gamma\geq\gamma$, we see from (\ref{As}) that $\beta$-order of $\v{r}$ is fixed to be (2,1,3). Moreover, the last elbow of $\beta(\v{r})$ has to lie on $\beta(\v{p})$, because the slope of a flattest segment ($\gamma$) is conserved by the transformation. In order to have the first elbow of $\beta(\v{r})$ on $\beta(\v{p})$, it is now enough that $q_{2}$ element is equal to $\partial p_{1} (q_{1,0}-q_{2,0})+p_{3} $. But this is exactly the value $a(q_{1,0}-q_{2,0})+c$ that is guaranteed by the transformation $A_{8}$ (\ref{pq}). 

By proceeding in the same way with all extremal points of TPs, we can verify that for each extremal TPs $A_{i}$ there exists a $\beta$-order such that for every $\v{p}$ with this $\beta$-order, $\v{r}=A_{i}\v{p}$ has $\beta$-order dependent only on $A_{i}$ and $\beta$-order of $\v{p}$, and all elbows of $\beta(\v{r})$ lie on $\beta(\v{p})$ (see Table \ref{Tab22}). Some curves formed by the action of chosen extermal TPs on a state of $\beta$-order (2,1,3) are shown in Fig. \ref{Fig3}.

Connection between $A_{9}$ and $\{A_{10},A_{11},A_{12},A_{13}\}$ is underlined by the fact that they transform states with the same order into each other (see Fig. \ref{Fig2}). The difference is that, in lower temperatures, the condition $q_{1,0}+q_{2,0}<1$ implies that two last segments of the state formed by the process maximizing slope on the first segment will be the same. This degeneration is reflected by the collapse of four extremal points $A_{10},A_{11},A_{12},A_{13}$ into a single one: $A_{9}$.

It remains an interesting question whether generalization of Prop. \ref{Prop3} holds. I.e., if for arbitrary $d$, every extremal TP $A$ can be matched to an initial state $\v{p}$ such that all elbows of $\beta(A\v{p})$ lie on $\beta(\v{p})$. If this was true, then it would be possible to calculate all extremal points of TPs for $d$ dimensional systems directly from thermomajorization diagrams. Namely, for a selected temperature $\beta$ it would be enough to investigate all thermomajorization curves with distinct and non-degeneratred $\beta$-order, for each curve determining the transformation that maps it to the curve with different $\beta$-order and whose all elbows lie on the initial curve. Every such a construction would be valid for a selected temperature range, therefore knowledge about values of threshold temperatures would be of a crucial importance. In the next section, we provide a construction determining the value of threshold temperatures for a given system Hamiltonian $H$.

\begin{figure}
\centering
\includegraphics[width = 1.0\linewidth]{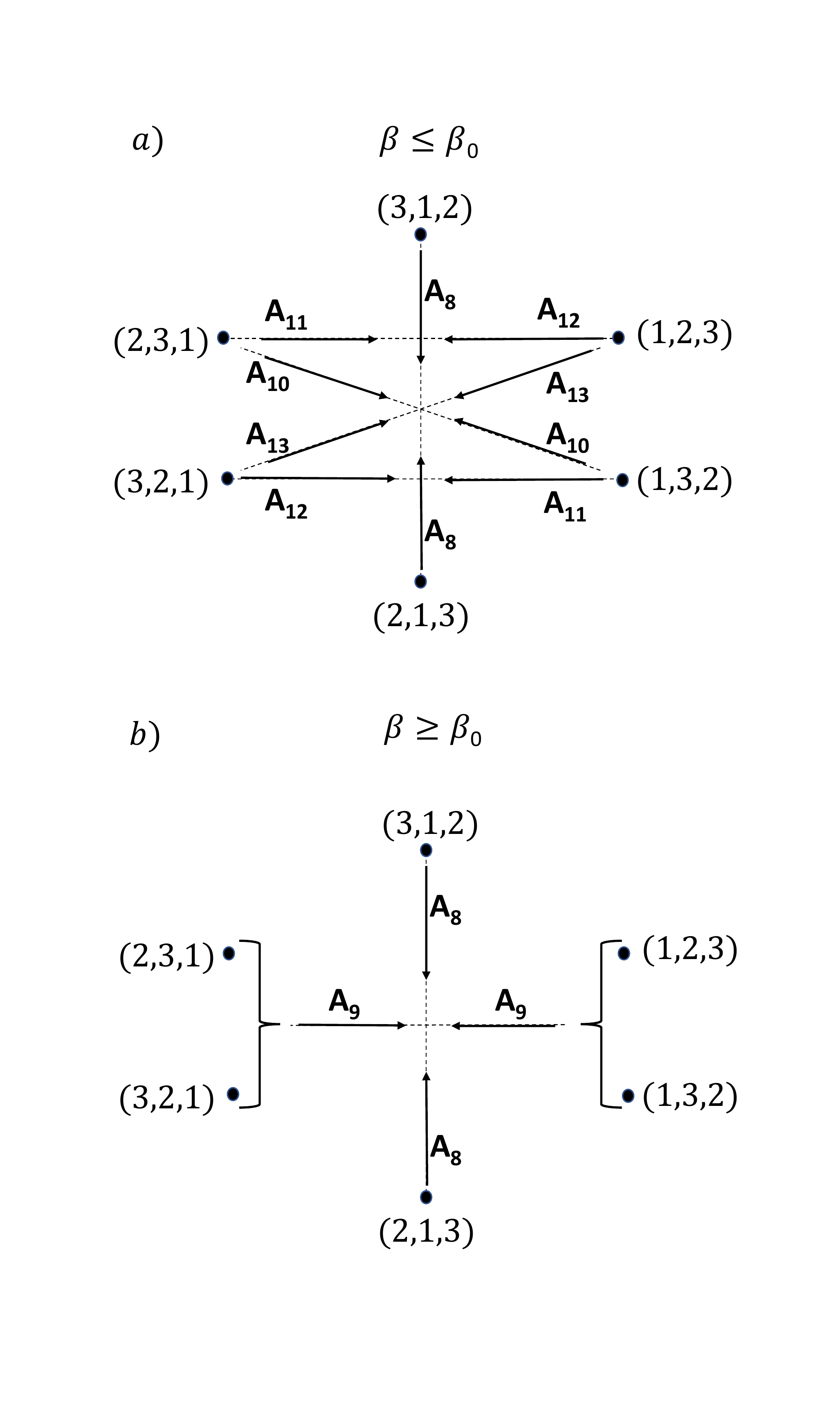}
\caption{\label{fig:sets} Mapping between states of given $\beta$-order provided by non-decomposable extremal points of TPs, for a) high and b) low temperatures. $A_{9}$ is low-temperature counterpart of $A_{10}, A_{11}, A_{12}, A_{13}$; slopes of last two segments of $\v{r}=A_{9}\v{p}$ are the same (for $\v{p}$ of the $\beta$-order presented in the picture). Braces mark a resulting degeneration of $\beta$-order of $\v{r}$. Connections between states provided by decomposable maps are not marked; they remain in agreement with Table I.} 
\label{Fig2}

\end{figure}

\subsection{Deterministic work extraction}
Here we would like to point out a connection between temperature dependence of the structure of the set of TPs and deterministically extractable work. Threshold temperatures 
that indicate change in the convex structure are clearly associated with relations between sums over components of a partition function: $\sum_{i\in A}q_{i,0}\geq\sum_{j\in B} q_{j,0}$, where $A$ and $B$ are disjoint set of indices, and $A, B\subset \{0,\dots,d-1\}$. Now, an incomplete sum of components of partition function is strictly related to 
min-free energy 
\begin{equation}
F_{min}(\v{p})=-kT \ln \sum_{i: p_{i}\neq 0} q_{i,0}.
\end{equation} 
introduced in \cite{Horodecki13} to describe the deterministically extractable work from a given state. 
The latter is given by $W_{extr}(\v{p})=F_{min}(\v{p})-F_{min}(\rho_{\beta})=-kT \ln \sum_{i: p_{i} \neq 0}q_{i,0}-(-kT\ln Z)$, 
where $Z$ is a partition function. Therefore, the order asserted by 
\begin{equation}
\label{eq:I}
\sum_{i\in A}q_{i,0}\geq\sum_{j\in B} q_{j,0}
\end{equation}
has an operational consequence as it determines the order among some states, in terms of work that can be extracted from them. Namely, 
\eqref{eq:I} is equivalent to 
\begin{equation}\label{w}
W_{extr}(\v{p}_{A}) \leq W_{extr}(\v{p}_{B}) 
\end{equation}
where the states $\v{p}_{A}$ and $\v{p}_{B}$ are arbitrary states which occupy solely levels belonging to $A$ and $B$, respectively. 
For example, the range of temperatures above the temperature $T_0$ of Eq. \eqref{eq:t_crit} is thus determined by the condition 
that the extractable work from ground state is greater than extractable work from state occupying second and third levels. 

TPs cannot lead to a transition which increases deterministically extractable work, as such a transition would violate the thermomajorozation condition (Prop. \ref{term}). Therefore, if there is an extremal TP that transforms states with occupations on $A$ set of levels to states with occupations on $B$ set of levels (with $A$ and $B$ being non-empty disjoint subsets of $\{0,\dots,d-1\}$), then we know that this TP cannot exist in the temperature regime in which $\sum\limits_{a\in A}q_{a,0}>\sum\limits_{b\in B}q_{b,0}$. On the other hand, for every pair of such disjoint sets $A$ and $B$ that admit $\sum\limits_{a\in A}q_{a,0}\leq\sum\limits_{b\in B}q_{b,0}$ for some temperature range one can always construct an extremal TP that transforms states occupying levels from the set $A$ to states occupying levels from set $B$ (we give the exact construction below). Therefore, if the sign of $\sum\limits_{a\in A}q_{a,0}-\sum\limits_{b\in B}q_{b,0}$ for a given Hamiltonian depends on temperature, then a system with this Hamiltonian admits the extremal TP only in the temperature range in which it would not violate the principle of non-increasing of deterministically extractable work. Hence we arrive at

\begin{prop}
For a system with a given Hamiltonian $H_{S}=\sum_{i=0}^{d-1}E_{i}|E_{i}\rangle\langle E_{i}|$ and for $q_{n,m}=e^{-\beta(E_{n}-E_{m})}$, $E_{0}=0$, every term $\sum\limits_{a\in A}q_{a,0}-\sum\limits_{b\in B}q_{b,0}$  with a sign depending on inverse temperature $\beta$, where $A,B$ are non-empty disjoint subsets of $\{0,\dots,d-1\}$, defines a threshold temperature, i.e. a temperature $\beta_{0}$: $\sum\limits_{a\in A}q_{a,0}=\sum\limits_{b\in B}q_{b,0}$ such that there is at least one extremal TP valid for $\beta\geq \beta_{0}$ and invalid for $\beta< \beta_{0}$, and at least one extremal TP valid for $\beta\leq \beta_{0}$ and invalid for $\beta> \beta_{0}$.
\end{prop}

{\textit{Construction of an extremal Thermal Processes associated with given threshold temperature.\\}}

Let us start with a term of the form $\sum\limits_{a\in A}q_{a,0}=\sum\limits_{b\in B}q_{b,0}$ from the Proposition above. Let us divide sets $A=\{n\}\bigcup I$, $B=\{m\}\bigcup J$ into subsets such that $n$ and $m$ are the smallest numbers from sets $A$ and $B$, respectively, and
 $I=\{i_{1},\dots,i_{|I|}\}$ and $J=\{j_{1},\dots,j_{|J|}\}$, and $i_{k}<i_{m}$ if $k<m$, the same for set $J$. Then, as long as 
$q_{n,0}+\sum\limits_{i\in I}q_{i,0}\geq  q_{m,0}+\sum\limits_{j\in J}q_{j,0}$, it is always possible to construct a TP of the form
\[
\begin{blockarray}{ccccccccc}
  & & m & & j_{1} & & & j_{|J|} & &\\
\begin{block}{c(cccccccc)}
 & \dots & 0 & \dots& 0& \dots& \dots& 0& \dots\\
n & \dots & 1-\sum\limits_{i\in I}q_{i,m} & \dots & 1 & \dots & \dots &1 & \dots\\
 & \dots & 0 & \dots& 0& \dots& \dots& 0& \dots\\
i_{1}& 0 & q_{i_{1},m} & 0& 0& 0& 0& 0& 0\\
& \dots & 0 & \dots& 0& \dots& \dots& 0& \dots\\
& \dots & 0 & \dots& 0& \dots& \dots& 0& \dots\\
i_{|I|} &0 & q_{i_{|I|},m} & 0& 0& 0& 0& 0& 0\\
& \dots & 0 & \dots& 0& \dots& \dots& 0& \dots\\
\end{block}
\end{blockarray}
 \]

This is because the Gibbs preserving condition applied to the $n$ row demands $(1-\sum\limits_{i\in I}q_{i,m})q_{m,0}+\sum\limits_{j\in J}1 q_{j,0}+y=q_{n,0}$, and, as long as $y=q_{n,0}-  q_{m,0}+\sum\limits_{i\in I}q_{i,0}-\sum\limits_{j\in J}q_{j,0}\geq 0$, one can always set the values of not-shown elements of the matrix such that the matrix is left stochastic and Gibbs preserving. This stems from the fact that every left stochastic and Gibbs preserving matrix, multiplied by a diagonal matrix $M_{\rho_{\beta}}$, can be turned into a transportation polytope \cite{Klee68} -- a matrix of non-negative elements with a property that elements of $k$ column and $l$ row sum to some number, $c_{k}$ and $r_{l}$, respectively . In our case, $r_{k}=c_{k}=q_{k,0}$. A set of transportation polytopes satisfying the given summation criteria is always non-empty as long as $\sum\limits_{k}c_{k}=\sum\limits_{k}r_{k}$. This is visible from the fact that, if $\sum\limits_{k}r_{k}=0$, the conditions are satisfied by a matrix with all elements equal to $0$. Otherwise, a matrix $A$ with elements $A_{i,j}=r_{i}c_{j}/\sum\limits_{k}r_{k}$ satisfies it. The existence of respective transportation polytopes is guaranteed also for a set of conditions that arises from fixing values of some matrix elements of the original matrix, as long as one fixes to 0 all other elements of the row(column) that the element was in, and subtracts the value of the fixed element from $c_{k}$ ($r_{k}$). This is exactly a process that describes fixing of shown matrix elements in the TP above. As there is a solution for the respective transportation polytope problem, there will be one as well for the case of the above left stochastic and Gibbs-preserving matrix. 

Note that the above TP maps all states with occupations on levels $\{m,i_{1},\dots,i_{|I|}\}$ into states with occupations on levels $\{n,j_{1},\dots,j_{|J|}\}$ -- a property that does not depend on the temperature. However, from (\ref{w}) we see that every process with such a property could lead to an increase of deterministic extractable work from a state whenever $q_{n,0}+\sum\limits_{i\in I}q_{i,0}<  q_{m,0}+\sum\limits_{j\in J}q_{j,0}$. Therefore, all processes with such a property, including the above process, have to cease at the temperature for which $q_{n,0}+\sum\limits_{i\in I}q_{i,0}=  q_{m,0}+\sum\limits_{j\in J}q_{j,0}$.

It is instructive to see that the above construction generates the appropriate extremal TPs for three level systems. There, we can have $A=\{0\}$ and $B=\{1,2\}$ under the assumption $q_{0,0}\geq q_{1,0}+q_{2,0}$. This leads to $n=0$, $m=1$ and $j_{1}=1$ and generates $A_{9}$ extremal TP. On the other hand, if one takes $A=\{1,2\}$ and $B=\{0\}$ under the assumption $q_{0,0}\leq q_{1,0}+q_{2,0}$, one gets $n=1$, $i_{1}=2$ and $m=0$, which leads to a TP described by a convex combination of extremal TPs $A_{11}$ and $A_{13}$. \\
The number of threshold temperatures depends on the Hamiltonian of the system. If we assume no degeneracies, then for $d$ level systems it is equal to the number of possible allocations of elements from the set $\{a_{1},a_{2},\dots,a_{d}\}$ with known order $a_{1}> a_{2}>\dots> a_{d}$ into two disjoint non-empty sets, such that the above order does not determine sum over elements from which set is bigger or equal to a sum over elements from the other set. Total number of possible allocations is given by $\frac{1}{2}\sum\limits_{k_{1}=1}\limits^{d-1}\sum\limits_{k_{2}=1}\limits^{d-k_{1}} \frac{d!}{k_{1}!k_{2}!(d-k_{1}-k_{2})!}=\frac{1}{2}(3^{d}-2^{d+1}+1)$, with a term under sums being number of possible different allocations of $k_{1}$ elements into first set and $k_{2}$ allocations into the second set, and a factor $\frac{1}{2}$ accounts for indistinguishability of the first and the second sets. Direct calculation of number of allocations satisfying the above criteria yields that the number of threshold temperatures for $d=3,4,5,6$ levels is equal to  $1,6,26,106$, respectively.

\section{Approximate transformations}\label{approx}
In Sec. \ref{transition} we gave an example of a transition that cannot be performed \textit{exacly} by TPs acting on 2 levels of the system: $\v{p}\xslashedrightarrowa{TP(2)} \v{r}$. A question arises about how the set of allowed transitions changes when we accept some error in the output state. Namely, we ask if for arbitrary $\epsilon>0$ there exists a state $\v{r}'$ such that: $|| \v{r}-\v{r'}||\leq \epsilon$ we have $\v{p}\xrightarrow{TP(2)} \v{r}'$. Below we show that for $\v{p}$ and $\v{r}$ taken from Sec. \ref{transition} such a state does not exist, i.e. there is some finite neighborhood of a state $\v{r}$ that TPs acting on 2 levels cannot lead to, and therefore they cannot be used to approximate $\v{r}$ from $\v{p}$ up to an arbitrary precision.

We will first sketch the idea of the proof for three level systems ($d=3$). 
An abitrary 2 level TP can be represented as a convex combination of sequences of extremal 2 level TPs. Let us start with investigating such sequences separately, and later generalise the result to the case of an arbitrary TP acting on two levels of the system.
 Since for two levels, there is just one extremal point (apart from identity), see eq. \eqref{2levels}, and there are three different pairs of levels, the sequence consists of one of three maps. 
One finds that for the chosen state, the map acting on two highest levels does not change the state. 
Hence, it is enough to consider sequences which start with one of the maps acting on levels $0$ and $1$ or $1$ and $2$
(denote them by $\Lambda_{0,1}$ and $\Lambda_{0,2}$).

Consider one of these maps, e.g. $\Lambda_{0,1}$ (for the other, the argument is the same). 
We shall now analyze the thermomajorization curve of the state $\v{r'}$ resulting from an arbitrary sequence starting with this map. 
Our aim will be to show, that such a curve will be bounded away from from the curve of the target state $\v{r}$.
This will be enough, because, if the curve of the state $\v{r'}$ cannot lie arbitrarily close to the curve of target state, 
then also the state $\v{r'}$ itself cannot lie arbitrarily close to the target state in statistical distance. 

\begin{figure}
\centering
\includegraphics[width = 1.0\linewidth]{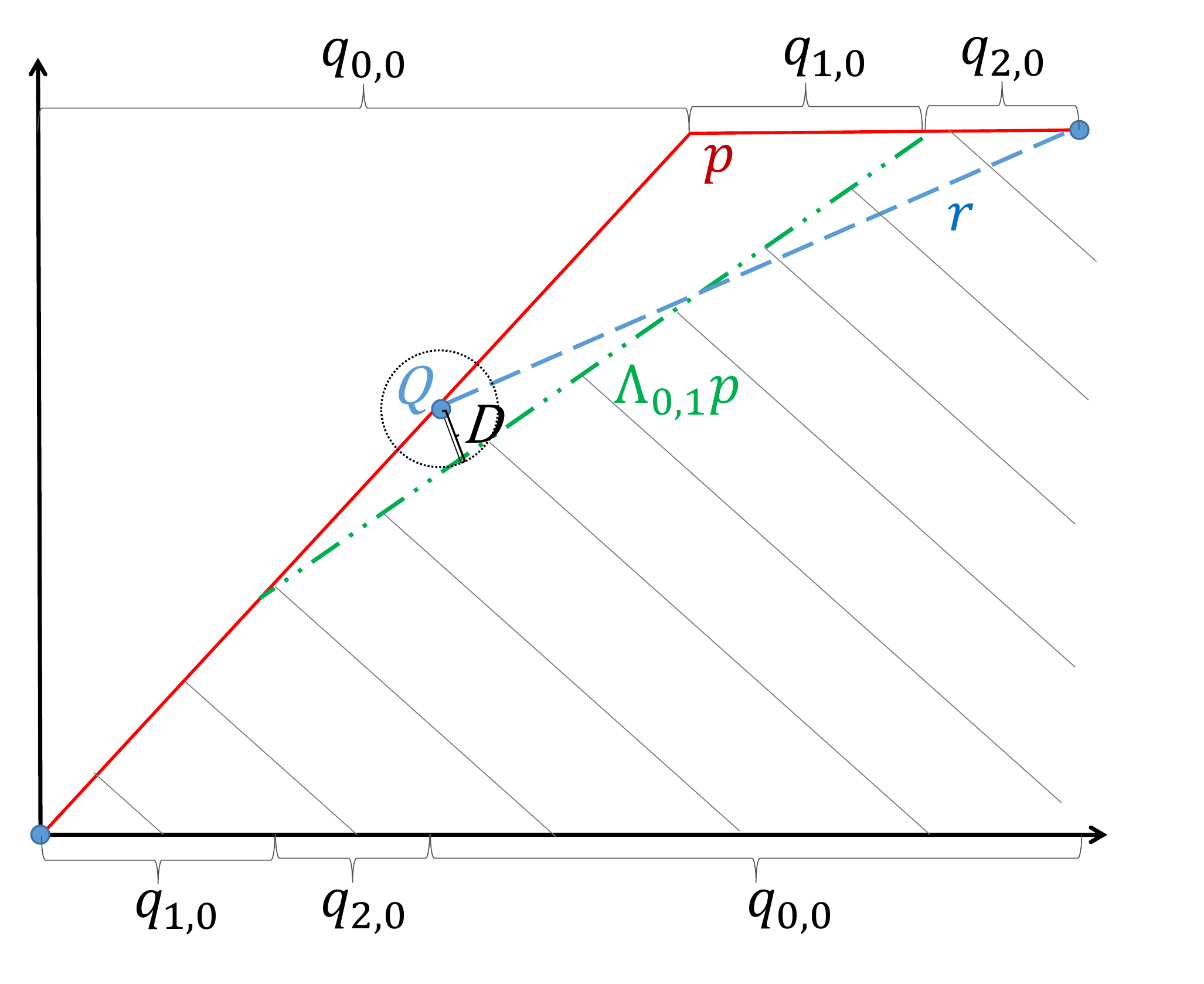}
\caption{\label{fig:approximate} Thermomajorization diagram for initial state $\v{p}$ (red solid curve), goal state $\v{r}$ (blue dashed curve) and state $\Lambda_{0,1}\v{p}$ (green dashed-dotted curve), emerging from $\v{p}$ after applying a 2 level extremal TP mixing levels $0$ and $1$. Through termomajorization condition it is visible that a curve of a state $\v{r}'=\Lambda_{rest}\Lambda_{0,1}\v{p}$, where $\Lambda_{rest}$ is a TP, can lie only within gray region, and therefore point $Q$ belonging to the curve of $\v{r}$ will be always separated from it at least by a distance $D>0$.}
\label{Fig2}

\end{figure}

Now, let us argue that curve of $\v{r'}$ must be indeed bounded away from that of $\v{r}$. Let us focus on the point $Q$ (see Fig. \ref{fig:approximate} ) on the curve of $\v{r}$.
After applying $\Lambda_{0,1}$ to $\v{p}$, it can be seen that the curve of the emereging state is bounded away from the curve of the target state $\v{r}$, as the separation $D$ between the point $Q$ and the curve $\Lambda_{0,1}\v{p}$ is always positive: $D>0$. Moreover, we see that subsequent application of another TP, call it $\Lambda_{rest}$, cannot lead to a curve of $\v{r}'=\Lambda_{rest}\Lambda_{0,1}\v{p}$ which converges with the curve of $\v{r}$: e.g., the point $Q$ on the curve of $\v{r}$ remains unattainable, and will be always separated from the curve of $\Lambda_{rest}\Lambda_{0,1}\v{p}$ at least by a distance $D>0$, set by the curve $\Lambda_{0,1}\v{p}$. This stems from the fact that every curve $\Lambda_{rest}\Lambda_{0,1}\v{p}$ lies no higher than the curve $\Lambda_{0,1}\v{p}$ due to thermomajorization condition (see Prop. \ref{term}).

Now, as thermomajorization curves of all states formed from $\v{p}$ by a sequence of 2 level TPs lie below the line of the target state, we see that convex combination of these sequences cannot make the thermomajorization line of the corresponding state approach the target line. Therefore, the transition cannot be performed up to an arbitrary precision by TPs acting on two levels.

Below we present a calculation of the lower bound of this minimal separation for arbitrary dimension $d$. We choose a metric $||\v{p}-\v{r}||=\sum\limits_{i}|p_{i}-r_{i}|$. The proof is based on the transformations of vectors which describe slopes of segments of given states on thermomajorization diagrams, as defined in Sec. \ref{thermo}. The relation $\partial x_{i}=x_{i}q_{0,i}$ for a given vector $\v{x}$ and its associated 'slope' vector $\partial\v{x}$, when applied to initial $\v{p}$ and final $\v{r}$ states, gives 

\begin{eqnarray}\label{slopes}
\partial\v{p}=
\begin{pmatrix}
1\\
0\\
0\\
\dots\\
0
\end{pmatrix},
&
\partial\v{r}=
\begin{pmatrix}
1-\sum\limits_{i=1}^{d-1}q_{i,0}\\
1\\
1\\
\dots\\
1
\end{pmatrix}.
\end{eqnarray}

As explained in Sec. \ref{thermo}, every Thermal Process $A$ such that $A\v{p}=\v{r}$ is associated with a map: $A^{s}\partial\v{p}=\partial\v{r}$ such that $A^{s}$ is a right stochastic matrix. In particular, every non-trivial, extremal TP on 2 different levels $k$ and $m$,  (see eq. \ref{2levels}), that we will denote $E(k,m)$, has the associated map $E^{s}(k,m)$ of the form 
\begin{equation}\label{A^{s}}
E^{s} (k,m)= \begin{pmatrix}
1-q_{m,k}& q_{m,k}  \\
1 & 0\\ 
\end{pmatrix}
\oplus Id,
\end{equation}
where $Id$ acts on the subspace of remaining levels. It implies that a slope of the higher level after transformation is equal to the slope of the lower level before the transformation, and the slope of the lower level is averaged. 

From the right-stochasticity of maps transforming slope vectors we see that, by performing a sequence of TPs, one cannot create a slope vector with increased maximal value. If we aim at obtaining a state $\v{r'}$ close to $\v{r}$, we have to apply some TPs connecting level $0$ with other levels, as this is the only way to obtain non-zero values of $\partial r_{j}$, $j=1,\dots,d-1$. Otherwise, $||\v{r}-\v{r'}||=2 \sum_{i\neq 0}q_{i,0}$. Therefore, we investigate possible impact which 2 level TPs applied to this state have on the distance. We concentrate on investigating sequences of extremal TPs, and show at the end, that allowing for mixed TPs cannot improve the distance. For the extremal case, based on the structure of $E^{s}(k,m)$, we conclude that the distance cannot be reduced to zero.

We have to start with some transformation $E(0,i)$, where $i=1,\dots,d-1$. We will describe cases $i=1$ and $i>1$ separately.\\

\underline{Case $i>1$.}
The following transformation of the initial slope vector takes place: 
\begin{eqnarray}\label{slopes}\nonumber
\begin{pmatrix}
1\\
0\\
\dots\\
0\\
0\\
0\\
\dots\\
0
\end{pmatrix}\xrightarrow{E^{s}(0,i), i>1}
&
\begin{pmatrix}
1-q_{i,0}\\
0\\
\dots\\
0\\
1\\
0\\
\dots\\
0
\end{pmatrix},
\end{eqnarray}
where $1$ in the output vector is at position $i$.
We see that further transformations are required, as at the moment we would have  $||\v{r}-\v{r}'||\geq |r_{i-1}-r_{i-1}'|= q_{i-1,0}>0$. Furthermore, we cannot leave an $i$ level untouched, as it would limit the achievable value $\partial r_{i-1} \leq 1-q_{i,0}\implies r_{i-1} = q_{i-1,0}(1-q_{i,0})\implies||\v{r}-\v{r}'||\geq |r_{i-1}-r_{i-1}'|= |q_{i-1,0}-q_{i-1,0}(1-q_{i,0})|= q_{i-1,0}q_{i,0}>0$. But performing a 2 level extremal TP on a level $i$ diminishes the maximal value present in the slope vector, with minimal reduction, to value $(1-q_{i,i-1})(1-q_{i,0})+q_{i,i-1}=1-q_{i,0}+q_{i,0}q_{i,i-1}$ happening for transformation between $i-1$ and $i$ levels, that follows after filling the level $i-1$ with the highest value possible:
\begin{eqnarray}\label{slopes2}\nonumber
\begin{pmatrix}
\text{up to } 1-q_{i,0}\\
\text{up to } 1-q_{i,0}\\
\dots\\
\text{up to } 1-q_{i,0}\\
1\\
\text{up to } 1-q_{i,0}\\
\dots\\
\text{up to } 1-q_{i,0}\\
\end{pmatrix}\xrightarrow{E^{s}(i-1,i)}
&
\begin{pmatrix}
\text{up to } 1-q_{i,0}\\
\text{up to } 1-q_{i,0}\\
\dots\\
\text{up to } 1-q_{i,0}+q_{i,0}q_{i,i-1}\\
\text{up to }1-q_{i,0}\\
\text{up to } 1-q_{i,0}\\
\dots\\
\text{up to } 1-q_{i,0}\\
\end{pmatrix},
\end{eqnarray}

Therefore, we have  $\partial r_{i-1} \leq 1-q_{i,0}+q_{i,0}q_{i,i-1}\implies r_{i-1} \leq q_{i-1,0}(1-q_{i,0}+q_{i,0}q_{i,i-1})\implies||\v{r}-\v{r}'||\geq |r_{i-1}-r_{i-1}'|= q_{i-1,0}^{2}(1-q_{i,i-1})>0$. Therefore, we see that by starting with $E(0,i)$ for $i>1$, we cannot approach state $\v{r}$ arbitrary close.\\

\underline{Case $i=1$.}
We start with the transformation

\begin{eqnarray}\label{slopes3}\nonumber
\begin{pmatrix}
1\\
0\\
0\\
\dots\\
\end{pmatrix}\xrightarrow{E^{s}(0,1)}
&
\begin{pmatrix}
1-q_{1,0}\\
1\\
0\\
\dots\\
\end{pmatrix},
\end{eqnarray}.
 
If no transformations that touch level $1$ followed, we would have $||\v{r}-\v{r}'||\geq |r_{2}-r_{2}'| = q_{2,0}q_{1,0}>0$. The following transformations cannot as well mix level $0$ with level $1$, as this would decrease the maximal slope present in the vector at least to  $(1-q_{1,0})^{2}+q_{1,0}$, and therefore would set a bound on the distance $||\v{r}-\v{r}'||\geq |r_{1}-r_{1}'| = q_{1,0}^{2}(1-q_{1,0})>0$. Therefore, the only option to increase $\partial r_{2}$ is to allow for some transformation connecting level $1$ with level $2$:

\begin{eqnarray}\label{slopes4}\nonumber
\begin{pmatrix}
a\\
1\\
b\\
c\\
\dots\\
\end{pmatrix}\xrightarrow{E^{s}(1,2)}
&
\begin{pmatrix}
a\\
(1-q_{2,1})b+q_{21}\\
1\\
c\\
\dots\\
\end{pmatrix},
\end{eqnarray}
where $a,b,c,\dots\geq 1-q_{1,0}$.

But if no transformation followed, this would set a bound on distance  $||\v{r}-\v{r}'||\geq |r_{1}-r_{1}'|= q_{1,0}-q_{1,0}((1-q_{2,1})b+q_{21})= q_{1,0}(q_{1,0}-q_{2,0})>0$. In order to decrease the distance, we have to increase the slope of $\partial r_{1}$, which can happen only by mixing levels $1$ and $2$, as all other levels have smaller slopes: $(1-q_{2,1})b+q_{21}\geq (1-q_{1,0})$. But it reduces the maximal slope present in the vector. It the same manner as in the case $i>1$, it can be shown that it leads to $\partial r_{1}\leq (1-q_{2,1})((1-q_{2,1})(1-q_{1,0})+q_{2,1})+q_{2,1}=1-2q_{2,1}-q_{1,0}(1-q_{2,1}^{2})$, which implies 
 $||\v{r}-\v{r}'||\geq |r_{1}-r_{1}'|= q_{1,0}-q_{1,0}(1-2q_{2,1}-q_{1,0}(1-q_{2,1}^{2}))= q_{1,0}(2q_{2,1}+q_{1,0}(1-q_{2,1})^{2})>0$. Therefore, starting with $E_{0,1}$, we cannot approach $\v{r}$ arbitrary close.\\

Therefore, by collecting all the bounds obtained above, we see that every sequence of 2 level extremal TPs applied to a state $\v{p}$ leads to a state $\v{r}'$ that satisfies $||\v{r}-\v{r}'||\geq \min\limits_{i>1}[q_{i-1,0}q_{i,0},q_{i-1,0}^{2}(1-q_{i,i-1}),q_{2,0}q_{1,0},q_{1,0}^{2}(1-q_{1,0}),q_{1,0}(q_{2,0}-q_{2,1}),q_{1,0}(2q_{2,1}+q_{1,0}(1-q_{2,1}^{2}))]>0$. Now it is enough to realise that performing a convex combination of arbitrary TPs is equivalent to performing a convex combination of sequences of extremal TPs. However, such a combination cannot lead to a state closer to $\v{r}$ than a state obtained by the most optimal of these sequences. This is because the bound on the distance calculated above relies on terms $|r_{i}-r_{i}'|$ for $i\geq 1$, and therefore, as in our case for all $\v{r'}$ that can be obtained from $\v{p}$ by 2 level extremal TPs, the value of $r_{i}-r_{i}'$ for $i\geq1$ is always non-negative, one cannot obtain reduction of the bound by allowing for $\v{r}'=\alpha\v{r}^{'(1)}+(1-\alpha)\v{r}^{'(2)}$ for $0\leq\alpha\leq 1$, where $\v{r}^{'(1)}$, $\v{r}^{'(2)}$ result from two sequences of extremal 2 level TPs.

\section{Discussion and conclusions}
We have presented a construction of Thermal Operation for arbitrary $d$ -level system, that cannot be performed without executing a joint operation on all energy levels. The extremal Thermal Process that performs the transformation exists for all temperatures low enough to allow for $\sum\limits_{i=1}^{d-1}e^{-\beta E_{i}}\leq 1$ to be satisfied. For three level systems, we have also identified counterpart processes for the remaining temperature range, showing their 
non-decomposability into a convex combination of composition of Thermal Processes acting non-trivially on 2 energy levels. We speculate that these processes can be generalized to an arbitrary dimension by exploiting the bipartite-graph structure associated with these 
matrices \cite{Klee68}. We also point out that some extremal points satisfy quantum detailed balance condition, whereas others form pairs with respect to conjugation according to an associated scalar product. The conjectured non-decomposibility of self-dual extremal points of Thermal Processes may be a helpful property in the analysis of the geometry of the set of $d$ level Thermal Processes.

One can try the solve the general decomposibility problem of Thermal Processes by analyzing the convex structure of the set, which probably would require determination of its extremal points. While pursuing the method of their computation that relies on fixing all matrix elements by some minimal number of zeros can be infeasible for larger $d$, exploitation of observed symmetries associated with quantum detailed balance condition and or/and gradual generation of extremal points of the set may lead to establishing a precise description of the geometry of the set of Thermal Processes that would take into account its decomposability into convex combination of products of more 'local' processes. In this, establishing a connection between the set of Thermal Processes and a set of all states possible to be obtained through Thermal Operations from a given initial state may be important. One should note e.g. that all states $\v{r}$ such that $\beta(\v{r})$ has all elbows on $\beta(\v{p})$ and is thermo-majorized by it, constitute all extremal points of this set 
\cite{Mazurek17}. Due to inability to increase the deterministically extractable work under Thermal Processes, in order to determine the full set of extremal points for systems with non-degenerated Hamiltonian it should be possible to focus on just two temperatures: one satisfying $1\geq\sum\limits_{i=1}^{d-1}e^{-\beta E_{i}}$, and the other $1\leq e^{-\beta E_{d-2}}+e^{-\beta E_{d-1}}$.

\textbf{Acknowledgments.} 
The results proving non-decomposability of some Thermal Processes into two levels (by means of different methods) have been independently obtained by M. Lostaglio, A. Alhambra and C. Perry in \cite{Lostaglio16b}. 
We would like to thank M. Lostaglio, A. Alhambra and C. Perry for inspiring discussions, and C. Perry in particular for drawing our attention to Ref. \cite{Gregory92}. M.H. and P. M. are supported by National Science Centre, Poland, grant OPUS 9. 2015/17/B/ST2/01945.

\bibliography{Extremal_biblio}

\end{document}